\documentclass[preprint,aps,superscriptaddress]{revtex4-1}
\usepackage{graphicx}         
\usepackage{dcolumn}         
\usepackage{bm}               
\usepackage{color}            
\usepackage{amsthm,amsmath,amssymb}  
\usepackage{amsfonts}       
\usepackage{multirow}      
\usepackage{verbatim}      
\usepackage{natbib}        
\usepackage{chemformula}
\usepackage{float}
\usepackage[colorlinks=true,linkcolor=green,anchorcolor=green,citecolor=green,urlcolor=green]{hyperref}

\begin{document}
	\title{Monolayer Fe$_{3}$GaX$_{2}$ (X=I, Br, Sb): high-temperature two-dimensional magnets and a novel partially ordered spin state}
	\author{Qiuhao Wang}
	\affiliation{School of Physics and Physical Engineering, Qufu Normal University, Qufu 273165, China.}
	\affiliation{Center for Quantum Physics, Key Laboratory of Advanced Optoelectronic Quantum Architecture and Measurement (MOE), School of Physics, Beijing Institute of Technology, Beijing 100081, China.}
	\affiliation{Beijing Key Lab of Nanophotonics $\&$ Ultrafine Optoelectronic Systems, School of Physics, Beijing Institute of Technology, Beijing 100081, China.}
	\author{Xinlong Yang}
	\affiliation{Center for Quantum Physics, Key Laboratory of Advanced Optoelectronic Quantum Architecture and Measurement (MOE), School of Physics, Beijing Institute of Technology, Beijing 100081, China.}
	\affiliation{Beijing Key Lab of Nanophotonics $\&$ Ultrafine Optoelectronic Systems, School of Physics, Beijing Institute of Technology, Beijing 100081, China.}
	\author{Fawei Zheng}
	\thanks{Corresponding author: fwzheng@bit.edu.cn}
	\affiliation{Center for Quantum Physics, Key Laboratory of Advanced Optoelectronic Quantum Architecture and Measurement (MOE), School of Physics, Beijing Institute of Technology, Beijing 100081, China.}
	\affiliation{Beijing Key Lab of Nanophotonics $\&$ Ultrafine Optoelectronic Systems, School of Physics, Beijing Institute of Technology, Beijing 100081, China.}
	\author{Ping Zhang}
	\thanks{Corresponding author: zhang\_ping@iapcm.ac.cn}
	\affiliation{School of Physics and Physical Engineering, Qufu Normal University, Qufu 273165, China.}
	\affiliation{Institute of Applied Physics and Computational Mathematics, Beijing 100088, China.}
	%\date{\today}
	\begin{abstract}
We systematically investigated the effects of charge doping and strain on monolayer Fe$_3$GaTe$_2$, and proposed three new novel two-dimensional magnetic materials: monolayer Fe$_3$GaX$_2$ (X=I, Br, Sb). We found that both strain and charge doping can tune the magnetic interactions, and the tuning by charge doping is more significant. Differential charge analysis revealed that the doped charges predominantly accumulate around Te atoms. Based on this insight, we introduced Fe$_{3}$GaI$_{2}$, Fe$_{3}$GaBr$_{2}$, and Fe$_{3}$GaSb$_{2}$ monolayers. The Fe$_{3}$GaI$_{2}$ and Fe$_{3}$GaBr$_{2}$ monolayers contain I and Br atoms rather than Te atoms, emulate electron-doped Fe$_{3}$GaTe$_{2}$ monolayer, resulting in notably high T$_c$ values of 867 K and 844 K, respectively. In contrast, the Fe$_3$GaSb$_2$ monolayer mimics  hole-doped Fe$_{3}$GaTe$_{2}$ monolayer, presents a mix of FM and antiferromagnetic interactions, manifesting a distinctive partially ordered magnetic state. Our study demonstrates that substitution atoms based on the charge-doping effect offer a promising approach for predicting new magnetic materials. The proposed Fe$_{3}$GaI$_{2}$, Fe$_{3}$GaBr$_{2}$, and Fe$_{3}$GaSb$_{2}$ monolayers hold great potential for spintronics applications, and may stimulate the pursuit of new types of spin liquid.
	\end{abstract}
	\maketitle
	In 2017, researchers reported two-dimensional (2D) materials with long-range ferromagnetic order\cite{1,2}. Since then, 2D magnets have attracted great attention because of their diverse physical properties and potential device applications. For example, combining a 2D magnet and 2D superconductor might produce a topological superconductor\cite{magsup} that can be used to construct a quantum computer. Combining two monolayer CrI$_3$ magnets leads to a bilayer CrI$_3$, which has stacking-dependent interlayer spin exchange interactions\cite{2018nano} that change the magnetic state from antiferromagnetic (AFM) to ferromagnetic (FM) and produces giant tunneling magnetoresistance\cite{3}. Furthermore, introducing a twist to the system can produce complex spin textures\cite{Xiao2021} and might support Skyrmions\cite{akram2021moire,6}, which can be used to construct memory devices and spin wave generators. These studies render 2D magnets promising materials for future nano-devices. However, these applications are limited by the Curie temperature (T$_{c}$) of 2D magnets. The T$_{c}$ of monolayer CrI$_{3}$ is 45 K\cite{1} and that of bilayer CrGeTe$_{3}$\cite{2,27}is even lower at only 28 K. Therefore, efforts have been made to increase the T$_{c}$ of 2D magnets. Tuning methods, such as strain\cite{11,26} and charge doping\cite{10,26}, can be used to increase the T$_c$ of 2D magnets. 
	The T$_c$ of CrI$_3$ can be increased by about 10\% under heavy charge doping\cite{7}, and increased by 36\% under 10\% stretch strain\cite{C5TC02840J}. In theoretical calculations, the T$_{c}$ of CrCl$_{3}$ and CrBr$_{3}$ is increased by 33\% and 60\%, respectively, under 10\% stretch strain\cite{C5TC02840J}, and the T$_c$ of CrGeTe$_{3}$ might increase by 45\% under charge doping\cite{Hou2023}. 

	In addition to tuning currently available 2D magnets, new 2D magnets have been found. First, atomically thin Fe$_{3}$GeTe$_{2}$ exhibits intrinsic  ferromagnetism, the T$_c$ of which is up to 200 K\cite{12,13}. In addition, monolayer Fe$_{5}$GeTe$_{2}$ \cite{Li2020} exhibits T$_c$ = 270 K. More examples can be found elsewhere\cite{15,CrSiTe}. Nevertheless, 2D magnets with a T$_{c}$ higher than room temperature remain rare. Recently, Fe$_{3}$GaTe$_{2}$ was reported\cite{14} to have a T$_{c}$ higher than 350 K. Its atomic structure is similar to that of Fe$_{3}$GeTe$_{2}$; the only difference is that the Ge atoms are replaced with Ga atoms. Further research\cite{20,22,23} indicates an excellent magnetoresistance of spin-valve devices composed of Fe$_{3}$GaTe$_{2}$, which function well at 300 K under low current conditions. In addition, the exchange bias effect of Fe$_{3}$GaTe$_{2}$ nanoflakes has been revealed experimentally\cite{24}. Even though many studies on Fe$_{3}$GaTe$_{2}$ have been reported, research on the fundamental properties of Fe$_{3}$GaTe$_{2}$ monolayers is still lacking.

	\begin{figure}
	\includegraphics[width=1.0\columnwidth]{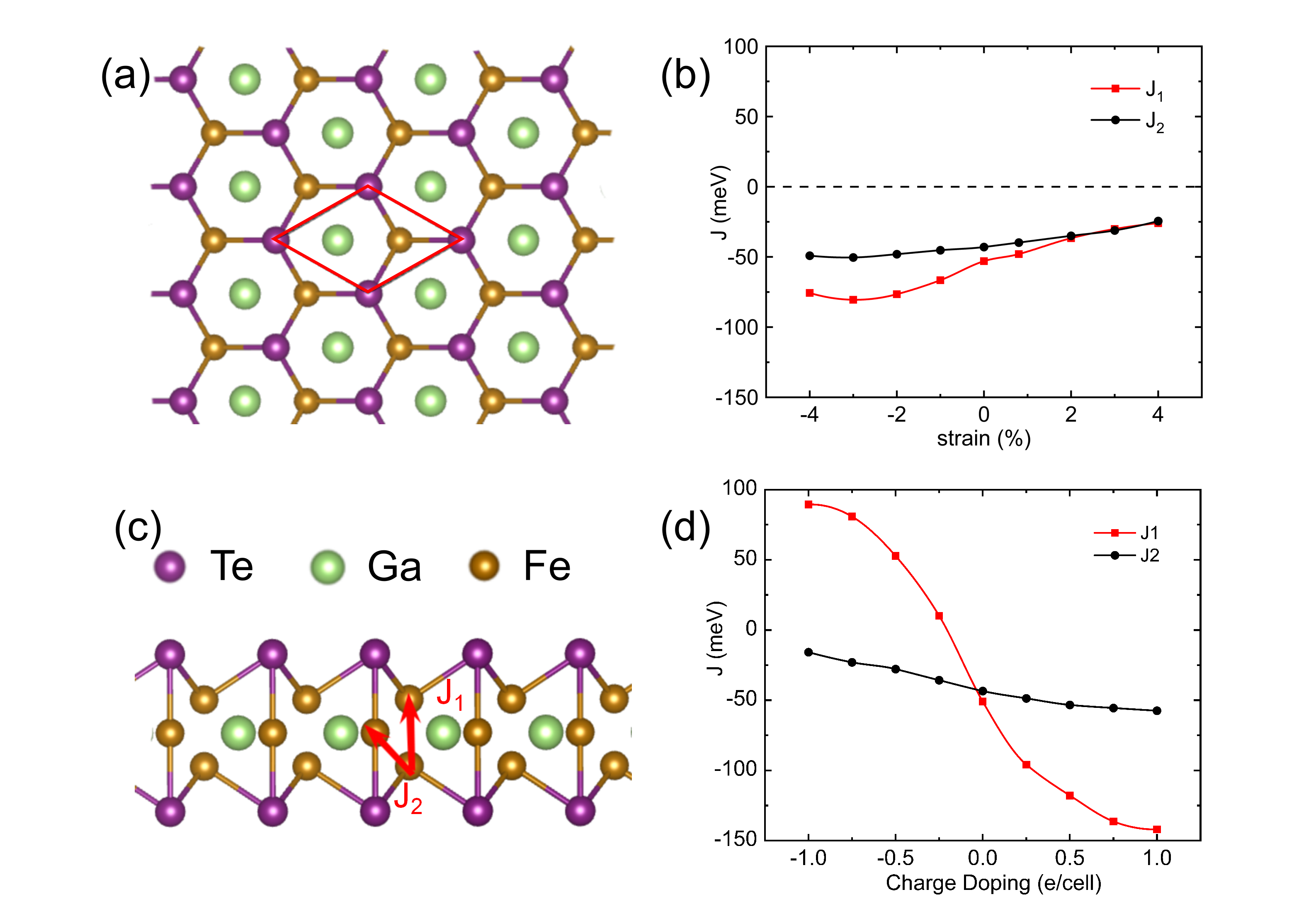}
	\caption{ The (a) top and (c) side views of an Fe$_{3}$GaTe$_{2}$ monolayer. The red rhombus in (a) indicates the primitive cell, and the two red arrows in (b) indicate the nearest and second-nearest neighboring spin exchange interactions. Values of the spin exchange constants under charge doping and biaxial strain are shown in (b) and (d), respectively.}
    \end{figure}

	In this work, on the basis of first-principles calculations and classical Monte Carlo simulations, we systematically studied the strain and charge doping effects of Fe$_{3}$GaTe$_{2}$ monolayers, and proposed three new 2D magnetic materials. We found that strain and charge doping are effective methods for tuning the T$_{c}$ of Fe$_{3}$GaTe$_{2}$ monolayers. The electron doping and compressive strain would increase the value of the T$_{c}$. The T$_c$ reaches 513 K under one electron doping per unit cell, and it is above 400 K under 2\% compressive strain. Conversely, hole doping and extensile strain weaken the spin exchange interactions, and hence decrease the T$_{c}$. On the basis of the detailed analysis of the doping charge distributions, we propose Fe$_3$GaI$_2$, Fe$_3$GaBr$_2$, and Fe$_3$GaSb$_2$ monolayer. Fe$_3$GaI$_2$ and Fe$_3$GaBr$_2$ have high T$_c$, and Fe$_{3}$GaSb$_{2}$ has a unique magnetic state that is partially ordered. 

	We performed all density functional theory calculations with the Vienna ab initio simulation package\cite{30,31}. To increase calculation efficiency, we used the projector augmented wave (PAW) method\cite{33} to describe the core electrons.  The exchange-correlation function was approximated using the generalized gradient approximation (GGA) parameterized by Perdew, Burke, and Ernzerhof (PBE)\cite{32}. The plane wave cutoff energy was set to 400 eV. Brillouin zone integrations were conducted on a 12×12×1 Gamma-centered K-point mesh. A vacuum of 20 \AA{} was employed to prevent interactions between periodic images. The cutoff energy, K-point mesh, and vacuum space were all optimized in our test calculations. The atomic structures were optimized until the force on each atom was less than 0.001 eV/\AA. Phonon dispersions were determined using the Phonopy package, where the real space force constants were obtained based on a 3×3×1 supercell and density functional perturbation theory (DFPT)\cite{34} as implemented in VASP. To obtain the value of Tc, the heat bath Monte Carlo method\cite{36} was employed. The simulations were performed on a 90×90×1 supercell with 150,000 simulation steps.
	
	\begin{figure*}
		\centering
		\includegraphics[width=0.8\linewidth]{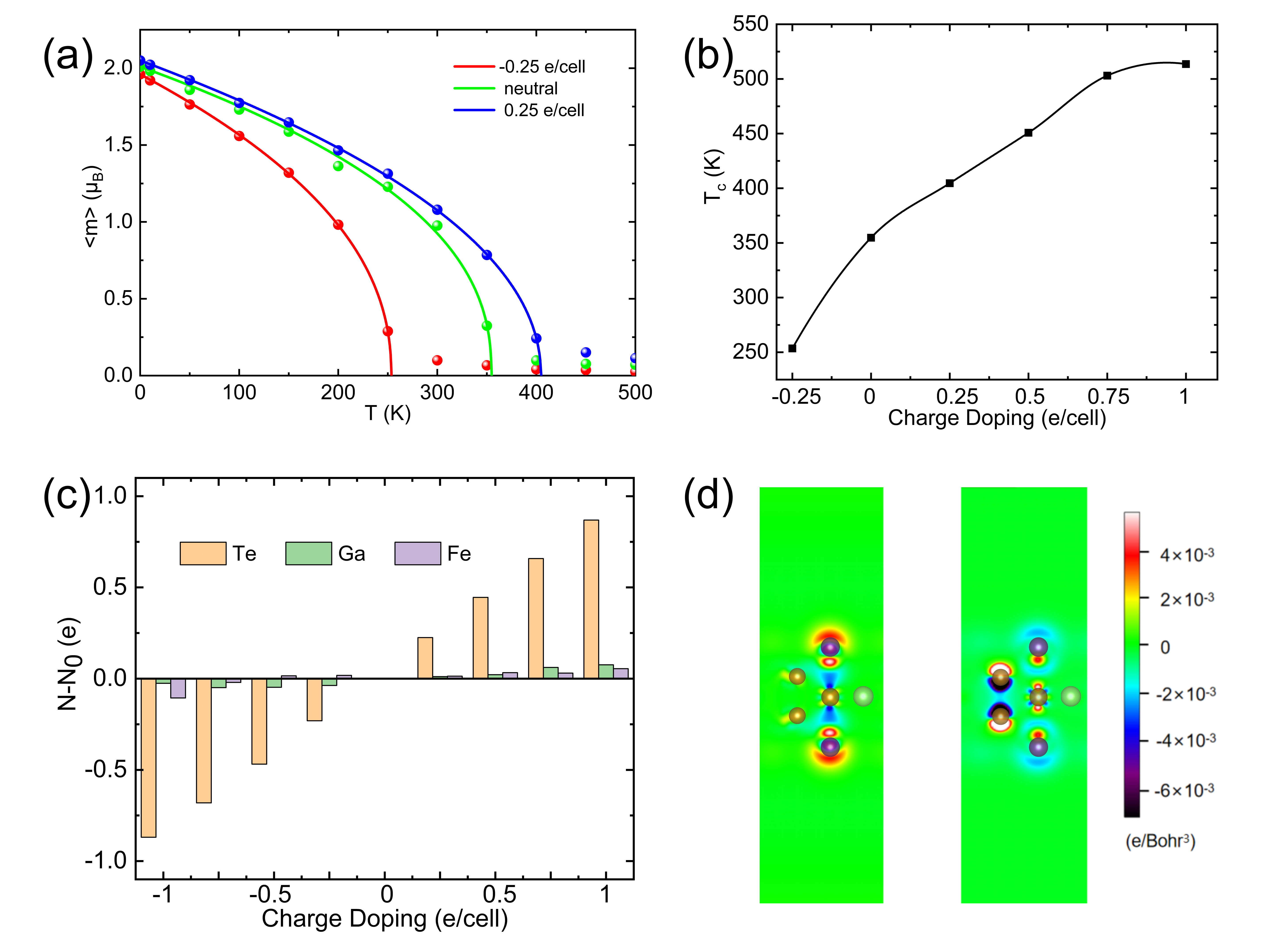}
		\caption{ (a) The magnetic polarizations as functions of temperature obtained by performing Monte Carlo simulations. (b) The curve for T$_c$ under various doping levels. (c) Doping charge accumulation for each element, obtained by using Bader's analysis. (d) Differential charge density for the 0.5-electron-doped (left panel) and 0.5-hole-doped (right panel) Fe$_3$GaTe$_2$ monolayer. }
	\end{figure*}

	The Figure 1a,c show the atomic structure of the Fe$_3$GaTe$_2$ monolayer. It consists of five layers of atoms. The top and bottom outermost layers consist of Te atoms, whereas the central layer consists of Ga and Fe atoms. The other two layers are all composed of Fe atoms. The magnetic moments of these Fe atoms differ. The magnetic moment of a Fe atom at the central layer is  1.4 $\mu_B$, whereas that of a Fe atom at the other layers is 2.3 $\mu_B$. All the Fe atoms form a hexagonal structure in the top view, and the Ga atoms are located at the centers of the hexagons. The relaxed lattice parameter is a=3.99\AA, which is close to the experiment value\cite{14}.

	The Fe$_3$GaTe$_{2}$ monolayer contains three layers of magnetic atoms; therefore, there are many different spin exchange interactions. The Hamiltonian of this system as,
	\begin{align*}
	H = \sum\limits_{ij} J_{ij} \bold{S}_i\cdot \bold{S}_j \approx J_{1}\sum\limits_{1nn} \bold{S}_i\cdot\bold{S}_j + J_{2}\sum_{2nn} \bold{S}_i\cdot\bold{S}_j
	\end{align*}
	where the $\bold{S}_i$ and $\bold{S}_j$ are spins at site i and j, and the constant J$_{ij}$ describes the spin exchange interaction between them. A previous study\cite{14} calculated these interactions up to the seventh nearest neighbors. The most important interactions are the first and second nearest neighboring interactions, as indicated in Figure 1a with $J_1$ and $J_2$. All the other interactions are much smaller than them. We checked (by performing Monte Carlo simulations) that the value of T$_c$ is determined by $J_1$ and $J_2$. The contribution from the other interactions is negligible. We obtained the values of $J_1$ and $J_2$ by using the four-state method\cite{35}. These values are -51.0 meV and -43.4 meV in our calculations, leading to T$_c$ = 355 K, which agrees well with the experimental results\cite{14}.
	
	\begin{figure*}[ht]
		\centering
		\includegraphics[width=0.8\linewidth]{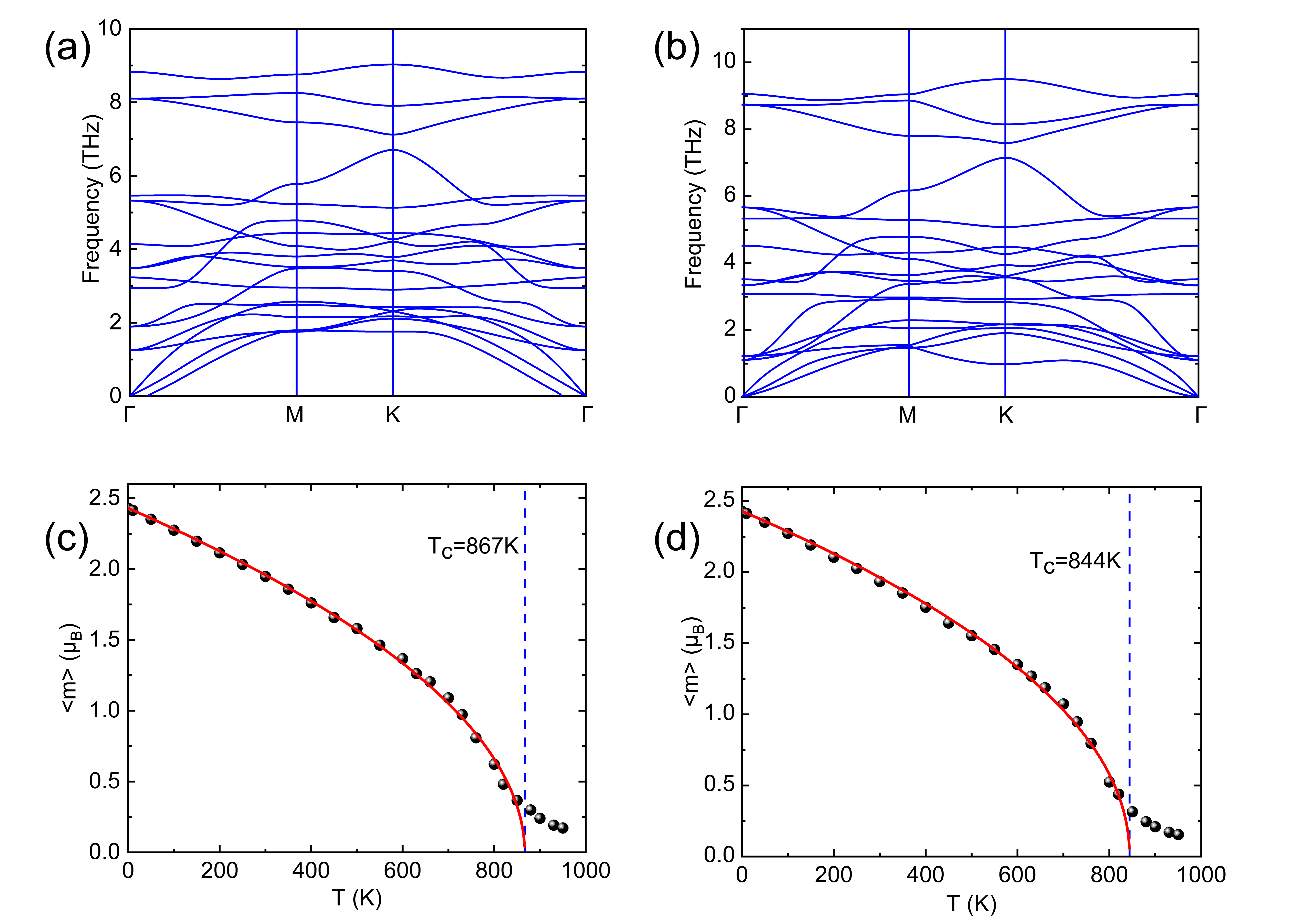}
		\caption{ The phonon dispersions of (quantuma) Fe$_3$GaI$_2$ and (b) Fe$_3$GaBr$_2$ monolayers. Their magnetic polarizations as functions of temperature are shown in (c) and (d), respectively.}
	\end{figure*}

	Firstly, we consider the strain effect of the Fe$_{3}$GaTe$_{2}$ monolayer. We applied biaxial strain in the range of $-4$\% $\sim$ 4\%. Figure 1b shows the calculated values of $J_1$ and $J_2$. The extensile strain monotonously weakens both $J_1$ and $J_2$. However, the compressive strain in the range of 0 $\sim$ $-3$\% enhances $J_1$ and $J_2$, thus increases the value of $T_c$. The values of $J_1$ and $J_2$ are $-76.6$ meV and $-48.1$ meV under $-2$\% compressive strain. The corresponding $T_c$ is above 400 K, which suggests that the strain is an effective method of enhancing $T_c$. Nevertheless, a compressive strain larger than $-3$\% weakens the $J_1$ interaction. 
	Another commonly used tuning method for 2D materials is charge doping. We thus studied the charge doping effect of the Fe$_{3}$GaTe$_{2}$ monolayer. The maximum doping charge density is 1.0 e/cell for electron doping, and $-1.0$ e/cell for hole doping. Figure 1c shows the calculated spin exchange constants. In the case of electron doping, the two FM interactions are enhanced. At 1.0 e/cell, the value of $J_1$ increases by 176\%, and that of $J_2$ increases by 32\%. In contrast, hole doping weakens the two FM interactions. The $J_1$ changes sign at a doping density of $-0.2$ e/cell, and becomes an AFM interaction at heavier hole doping. Compared with the strain effect, the charge doping effect changes the $J_1$ and $J_2$ more substantially, rendering it a more effective means of tuning the magnetic properties of the Fe$_{3}$GaTe$_{2}$ monolayer.
	
	We also calculated the tuning of T$_c$ under charge doping. Figure 2a shows the simulated mean moments as functions of temperature. We calculated the values of $T_c$ by fitting these data with $<m>=m_0(1-T/T_c)^\beta$, where $m_0=1$ for our Hamiltonian with normalized spins. The fitting result indicates that the $T_c$ without charge doping is 355 K. The $T_c$ increases to 404 K at 0.25 e/cell electron doping, and decreases to 253 K at $-0.25$ e/cell hole doping. At $-0.25$ e/cell hole doping, $J_1$ has a small positive value, indicative of a weak AFM interaction. However, $J_2$ has a large negative value, which dominates the magnetic properties of the system. Thus, the system is still a FM, and the $T_c$ is decreased by  30\%.

	To obtain the doping charge distribution, we performed Bader's analysis calculations\cite{bader}. We used the charge on each element after doping minus that of the neutral case to indicate the distribution of the doping charge. Figure 2c shows the calculation results, in which the charge on an element is the summation of the doping charges on all the atoms of the element in a unit cell. A large quantity of the doping charges is located at the Te atoms. When the doping is 1 e/cell, the Te atoms absorb 85\% of the total doping electrons. Further calculations indicate that this ratio is independent of the doping level. This conclusion is also confirmed by the calculation of differential charge density as shown in Figure 2d, which is defined as the difference of charge density after and before doping. The differential charge density around the Ga atom is small, which explains the small doping charges in Bader's analysis. The differential charge density on the central Fe atom differs from that of the other two Fe atoms. Their values cancel, leaving a small doping charge on the Fe element. The most doping charges are around the Te atoms, located at the outside of the surface Te atoms. 
	\begin{figure*}
		\centering
		\includegraphics[width=0.8\linewidth]{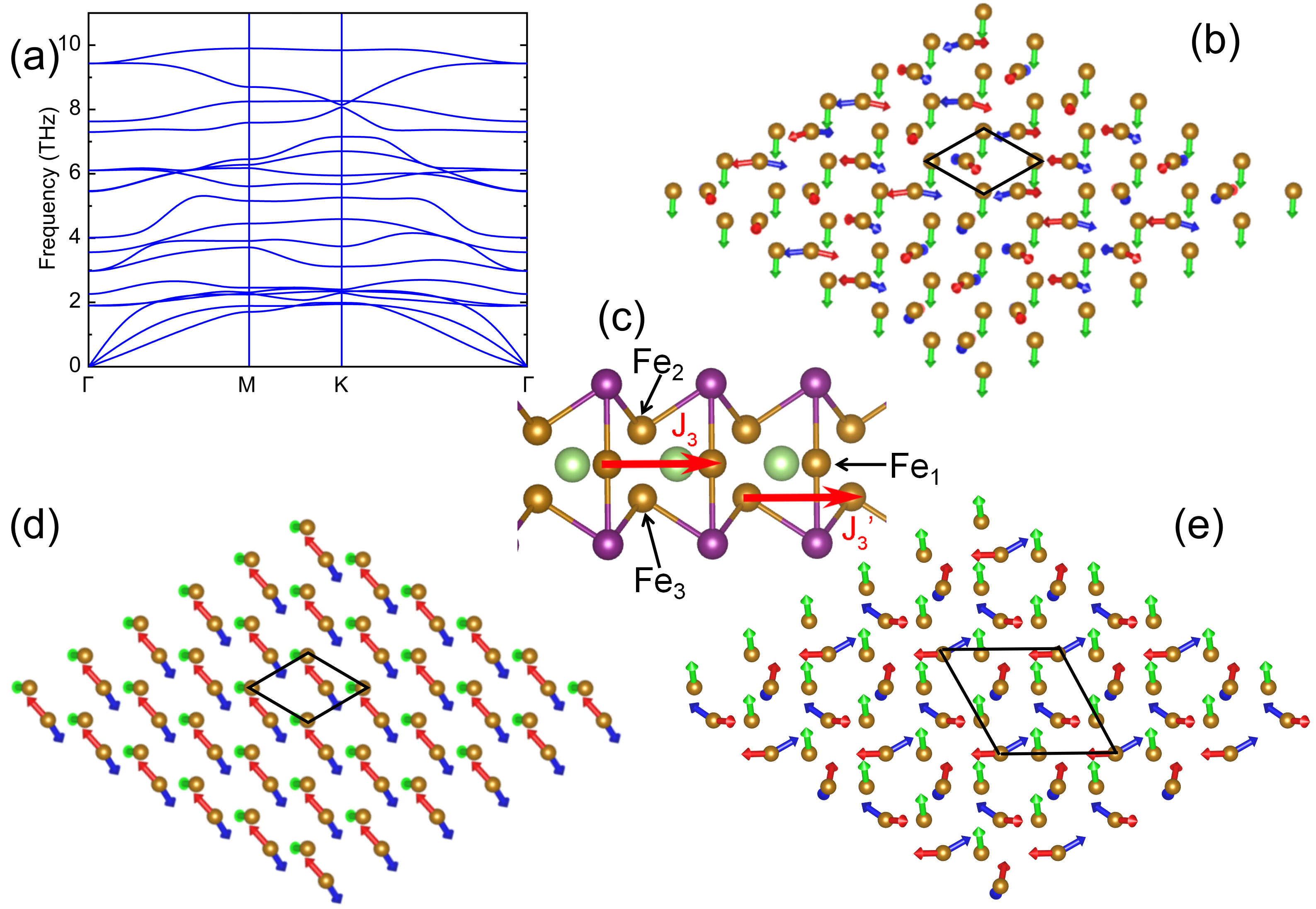}
		\caption{(a) Phonon dispersions of Fe$_{3}$GaSb$_{2}$ monolayer. (b) Partially ordered ground magnetic state of Fe$_{3}$GaSb$_{2}$ monolayer. (c) Red arrows indicate the third nearest neighboring spin exchange interactions J$_3$ and J$_3$'. (d) Ground state after considering the weak interactions J$_3$ and J$_3$'. (e) Ground state of Fe$_{3}$GaSb$_{2}$ monolayer under 2\% extensile strain.}
	\end{figure*}
	Almost all the doping charges are absorbed by the Te atoms, and therefore it is natural to simulate the heavy charge doping by substituting Te with other elements. For example, I and Br have one more valence electron than Te, and thus substituting Te would simulate heavy electron doping. Sb has one less valence electron than Te and thus can be used to simulate heavy hole doping.
	 
	After substituting Te atoms with I and Br atoms, two new materials are obtained: Fe$_{3}$GaI$_{2}$ and Fe$_{3}$GaBr$_{2}$ monolayers. Their lattice parameters are 4.13 and 4.10\AA\, which are slightly larger than that of the Fe$_{3}$GaTe$_{2}$ monolayer. We checked their stabilities by the phonon dispersions. The Figure 3a,b show the calculated phonon dispersions based on density functional perturbation theory, which clearly indicate that they are stable. First-principles molecular dynamics simulations also confirm that Fe$_{3}$GaI$_{2}$ and Fe$_{3}$GaBr$_{2}$ monolayers are stable at room temperature. 	
	In monolayer Fe$_{3}$GaI$_{2}$ and Fe$_{3}$GaBr$_{2}$, the magnetic moment of the Fe atoms at the central layer are 2.1 $\mu_B$, and those of the other layers are 2.6 $\mu_B$. They are larger than that in the Fe$_{3}$GaTe$_{2}$ monolayer. By using the four-state method, we obtained the spin exchange constants of Fe$_{3}$GaI$_{2}$ and Fe$_{3}$GaBr$_{2}$ monolayers. The value of $J_1$ for the Fe$_{3}$GaI$_{2}$ monolayer is $-196.8$ meV, which is about 3 times larger than that of the Fe$_{3}$GaTe$_{2}$ monolayer. The value of $J_2$ is $-97.4$ meV. This is almost 2.2 times of the value in the Fe$_{3}$GaTe$_{2}$ monolayer. The values for the Fe$_{3}$GaBr$_{2}$ monolayer are $J_1$ = $-170.3$ and $J_2$ = $-99.7$ meV. These values are slightly smaller than that of the Fe$_{3}$GaI$_{2}$ monolayer, but still much larger than that of the Fe$_{3}$GaTe$_{2}$ monolayer. These trends agree with the charge doping effect in the Fe$_{3}$GaTe$_{2}$ monolayer. These strong FM interactions lead to high phase transition temperatures. The Figure 3c,d show the calculated curves of magnetization. The fitted transition temperatures are T$_{c}$ = 867 K for the Fe$_{3}$GaI$_{2}$ monolayer and T$_{c}$ = 844 K for the Fe$_{3}$GaBr$_{2}$ monolayer. They are much larger than room temperature, imparting these two materials with very stable FM states at room temperature. 
	
	Substitution of Te atoms in an Fe$_{3}$GaTe$_{2}$ monolayer with Sb atoms leads to a new material: an Fe$_{3}$GaSb$_{2}$ monolayer. An Sb atom has an approximately equal radius as a Te atom, and has one less valence electron. Thus, the substitution mimics the heavy hole doping of the Fe$_{3}$GaTe$_{2}$ monolayer. The phonon dispersions as shown in Figure 4a have no imaginary frequencies, suggesting that the system is stable. First-principles molecular dynamics simulations confirmed that the system is stable at room temperature. Details are in the Supporting Information.
	The magnetic moment of the central Fe atom (Fe$_1$) in the Fe$_{3}$GaSb$_{2}$ monolayer [Figure 4b] is 1.1 $\mu_B$, and the magnetic moments for Fe$_2$ and Fe$_3$ are all 2.0 $\mu_B$. These values are substantially smaller than that of the Fe atoms in the Fe$_{3}$GaTe$_{2}$ monolayer.
	
    We calculated the spin exchange constants between these magnetic atoms. The value of $J_1$ for the Fe$_{3}$GaSb$_{2}$ monolayer is 78.2 meV, which is AFM rather than FM. The value of $J_2$ is $-8.2$ meV, which is FM but is much weaker than that of the neutral Fe$_{3}$GaTe$_{2}$ monolayer. Their trends are consistent with the hole doped Fe$_{3}$GaTe$_{2}$ monolayer as shown in Figure 1d.
	The strength of the AFM interaction described by $J_1$ is one order of magnitude larger than that of the FM interaction described by $J_2$. Therefore, each pair of Fe$_2$ and Fe$_3$ atoms always has antiparallel magnetic moments (normalized moments are named as $\bold{m}_2$ and $\bold{m}_3$). For each pair of Fe$_2$-Fe$_3$, there are three nearest Fe$_1$ atoms. The couplings between Fe$_1$ and Fe$_2$-Fe$_3$ are FM. Thus, the magnetic moments $\bold{m}_2$ and $\bold{m}_3$ are slightly tilted to the direction of the magnetic moment on Fe$_1$ ($\bold{m}_1$). Such tilting further polarizes the other two Fe$_1$ atoms. Thus, all the Fe$_1$ atoms have a parallel magnetic moment at the ground state. Suppose the Fe$_1$ atoms have all magnetic moment in $x$-direction, then the energy raised by a Fe$_2$ and Fe$_3$ can be written as: 
	\begin{align*}
	E_{23} &= J_{1}\bold{m}_2\cdot \bold{m}_3+3J_{2}\bold{m}_1\cdot \bold{m}_3+3J_{2}\bold{m}_1\cdot \bold{m}_2\\
		&= \frac{1}{2}J_{1}(\bold{m}_2+\bold{m}_3)^{2}+3J_{2}\bold{m}_1\cdot (\bold{m}_2+\bold{m}_3)+c.
	\end{align*}
    The first term is determined by the norm of $\bold{m}_2+\bold{m}_3$, the second term is the projection of $\bold{m}_2+\bold{m}_3$ to $x$-direction, and the last term is a constant $c=-J_1\frac{m_2^2+m_3^2}{2}$. Since the $J_2$ is negative, the minimum energy could be obtained by vary the direction of $\bold{m}_2+\bold{m}_3$ to the $x$-direction, then the $E_{23}= \frac{1}{2}J_{1}|\bold{m}_2+\bold{m}_3|^{2}+3J_{2}|\bold{m}_2+\bold{m}_3|+c$.  Then the $\bold{m}_2+\bold{m}_3$ in the ground state can be obtained by 
    $\frac{\partial E_{23}}{\partial|\bold{m}_2+\bold{m}_3|}=0$. The results show that the angle between $\bold{m}_2$ and $\bold{m}_3$ is 162$^\circ$.
    We can see that the energy only depends on $\bold{m}_2+\bold{m}_3$, thus it leaves the $\bold{m}_2-\bold{m}_3$ an unstrained freedom. There should be infinite number of ground states in which each pair of Fe$_2$-Fe$_3$ atoms have a randomly chosen direction of $\bold{m}_2$-$\bold{m}_3$. This analysis is confirmed by our Monte Carlo simulations, which provide the ground state as a peculiar partially ordered state, as shown in Figure 4b. Further weak interactions between Fe$_2$-Fe$_3$ pairs might remove the randomness, and lead to a fully ordered ground state. 
    
    We then considered the third nearest interactions $J_3$ and $J_3'$ as shown in Figure 4c. The calculated $J_{3}$ is $-2.7$ meV and $J_3'$ is $-0.6$ meV. These FM interactions force the magnetic moments of Fe$_2$--Fe$_3$ pairs to be distributed in an ordered manner as shown in Figure 4d. A small biaxial tensile strain (2\%) can change the sign of $J_3'$, which brings an ordered state in the $\sqrt{3}\times\sqrt{3}$ supercell as shown in Figure 4e. The value of $J_3'$ is 0.7 meV under biaxial tensile strain. Therefore, the Fe$_{3}$GaSb$_{2}$ monolayer has a complex ground magnetic state and is sensitive to the external biaxial tensile strain.  The ordered states in Figure 4d,e might be easily destroyed by small thermal or quantum fluctuations, and become partially ordered states again.

	In conclusion, we used first-principles calculations and Monte Carlo simulations to investigate the magnetic interactions in Fe$_{3}$GaTe$_{2}$ monolayers under charge doping and biaxial strain, which leads to the propose of three novel 2D magnetic materials. Both the strain and charge doping can tune the magnetic interactions of the Fe$_{3}$GaTe$_{2}$ monolayer. A higher T$_{c}$ can be obtained by electron doping or compressive strain, and the charge doping is more efficient than biaxial strain. The sign of $J_1$ can even be changed under heavy hole doping. By using the differential charge density and Bader's analysis, we reveal that the doped charge is mainly distributed around the Te atoms. Therefore, we propose three materials to mimic heavy electron doping and hole doping of Fe$_{3}$GaTe$_{2}$ monolayers: Fe$_{3}$GaI$_{2}$, Fe$_{3}$GaBr$_{2}$, and Fe$_{3}$GaSb$_{2}$ monolayers. I and Br atom substitution leads to Fe$_{3}$GaI$_{2}$ and Fe$_{3}$GaBr$_{2}$ monolayers, which are analogous to the heavy electron doped Fe$_{3}$GaTe$_{2}$ monolayer, have strong FM interactions, and lead to astonishingly high T$_c$. On the other hand, Sb atom substitution leads to an Fe$_{3}$GaSb$_{2}$ monolayer, which is analogous to the heavy hole doped Fe$_{3}$GaTe$_{2}$ monolayer. It has a unique partially ordered magnetic state. The model Hamiltonian and the special magnetic state inspires the possible existence of the partially ordered spin liquid. Our study provides an effective method to find new 2D magnetic materials; the proposed Fe$_{3}$GaI$_{2}$ and Fe$_{3}$GaBr$_{2}$ monolayers have potential applications in room-temperature spintronics, and the novel magnetic state of Fe$_{3}$GaSb$_{2}$ monolayer may stimulate the study of new spin liquids.

	This work was supported by the National Natural Science Foundation of China under Grant Nos. 12022415, 12374054, and 11974056. We acknowledge the computing resources of the Tencent TEFS platform (https://tefscloud.com). 

\providecommand{\latin}[1]{#1}
\makeatletter
\providecommand{\doi}
  {\begingroup\let\do\@makeother\dospecials
  \catcode`\{=1 \catcode`\}=2 \doi@aux}
\providecommand{\doi@aux}[1]{\endgroup\texttt{#1}}
\makeatother
\providecommand*\mcitethebibliography{\thebibliography}
\csname @ifundefined\endcsname{endmcitethebibliography}
  {\let\endmcitethebibliography\endthebibliography}{}
	

\begin{mcitethebibliography}{33}
\providecommand*\natexlab[1]{#1}
\providecommand*\mciteSetBstSublistMode[1]{}
\providecommand*\mciteSetBstMaxWidthForm[2]{}
\providecommand*\mciteBstWouldAddEndPuncttrue
  {\def\EndOfBibitem{\unskip.}}
\providecommand*\mciteBstWouldAddEndPunctfalse
  {\let\EndOfBibitem\relax}
\providecommand*\mciteSetBstMidEndSepPunct[3]{}
\providecommand*\mciteSetBstSublistLabelBeginEnd[3]{}
\providecommand*\EndOfBibitem{}
\mciteSetBstSublistMode{f}
\mciteSetBstMaxWidthForm{subitem}{(\alph{mcitesubitemcount})}
\mciteSetBstSublistLabelBeginEnd
  {\mcitemaxwidthsubitemform\space}
  {\relax}
  {\relax}

\bibitem[Huang \latin{et~al.}(2017)Huang, Clark, Navarro-Moratalla, Klein,
  Cheng, Seyler, Zhong, Schmidgall, McGuire, Cobden, Yao, Xiao,
  Jarillo-Herrero, and Xu]{1}
Huang,~B.; Clark,~G.; Navarro-Moratalla,~E.; Klein,~D.~R.; Cheng,~R.;
  Seyler,~K.~L.; Zhong,~D.; Schmidgall,~E.; McGuire,~M.~A.; Cobden,~D.~H.;
  Yao,~W.; Xiao,~D.; Jarillo-Herrero,~P.; Xu,~X. Layer-dependent ferromagnetism
  in a van der Waals crystal down to the monolayer limit. \emph{Nature}
  \textbf{2017}, \emph{546}, 270--273\relax
\mciteBstWouldAddEndPuncttrue
\mciteSetBstMidEndSepPunct{\mcitedefaultmidpunct}
{\mcitedefaultendpunct}{\mcitedefaultseppunct}\relax
\EndOfBibitem
\bibitem[Marti \latin{et~al.}(2014)Marti, Fina, Frontera, Liu, Wadley, He,
  Paull, Clarkson, Kudrnovsk{\'y}, Turek, Kune{\v{s}}, Yi, Chu, Nelson, You,
  Arenholz, Salahuddin, Fontcuberta, Jungwirth, and Ramesh]{2}
Marti,~X. \latin{et~al.}  Room-temperature antiferromagnetic memory resistor.
  \emph{Nature Materials} \textbf{2014}, \emph{13}, 367--374\relax
\mciteBstWouldAddEndPuncttrue
\mciteSetBstMidEndSepPunct{\mcitedefaultmidpunct}
{\mcitedefaultendpunct}{\mcitedefaultseppunct}\relax
\EndOfBibitem
\bibitem[Kezilebieke \latin{et~al.}(2020)Kezilebieke, Huda, Va{\v{n}}o, Aapro,
  Ganguli, Silveira, G{\l}odzik, Foster, Ojanen, and Liljeroth]{magsup}
Kezilebieke,~S.; Huda,~M.~N.; Va{\v{n}}o,~V.; Aapro,~M.; Ganguli,~S.~C.;
  Silveira,~O.~J.; G{\l}odzik,~S.; Foster,~A.~S.; Ojanen,~T.; Liljeroth,~P.
  Topological superconductivity in a van der Waals heterostructure.
  \emph{Nature} \textbf{2020}, \emph{588}, 424--428\relax
\mciteBstWouldAddEndPuncttrue
\mciteSetBstMidEndSepPunct{\mcitedefaultmidpunct}
{\mcitedefaultendpunct}{\mcitedefaultseppunct}\relax
\EndOfBibitem
\bibitem[Sivadas \latin{et~al.}(2018)Sivadas, Okamoto, Xu, Fennie, and
  Xiao]{2018nano}
Sivadas,~N.; Okamoto,~S.; Xu,~X.; Fennie,~C.~J.; Xiao,~D. Stacking-Dependent
  Magnetism in Bilayer CrI$_3$. \emph{Nano Letters} \textbf{2018}, \emph{18},
  7658--7664\relax
\mciteBstWouldAddEndPuncttrue
\mciteSetBstMidEndSepPunct{\mcitedefaultmidpunct}
{\mcitedefaultendpunct}{\mcitedefaultseppunct}\relax
\EndOfBibitem
\bibitem[Song \latin{et~al.}(2018)Song, Cai, Tu, Zhang, Huang, Wilson, Seyler,
  Zhu, Taniguchi, Watanabe, McGuire, Cobden, Xiao, Yao, and Xu]{3}
Song,~T.; Cai,~X.; Tu,~M. W.-Y.; Zhang,~X.; Huang,~B.; Wilson,~N.~P.;
  Seyler,~K.~L.; Zhu,~L.; Taniguchi,~T.; Watanabe,~K.; McGuire,~M.~A.;
  Cobden,~D.~H.; Xiao,~D.; Yao,~W.; Xu,~X. Giant tunneling magnetoresistance in
  spin-filter van der Waals heterostructures. \emph{Science} \textbf{2018},
  \emph{360}, 1214--1218\relax
\mciteBstWouldAddEndPuncttrue
\mciteSetBstMidEndSepPunct{\mcitedefaultmidpunct}
{\mcitedefaultendpunct}{\mcitedefaultseppunct}\relax
\EndOfBibitem
\bibitem[Xiao \latin{et~al.}(2021)Xiao, Chen, and Tong]{Xiao2021}
Xiao,~F.; Chen,~K.; Tong,~Q. Magnetization textures in twisted bilayer
  $\mathrm{Cr}{\mathrm{X}}_{3}$ ($\mathrm{X}$=Br, I). \emph{Phys. Rev. Res.}
  \textbf{2021}, \emph{3}, 013027\relax
\mciteBstWouldAddEndPuncttrue
\mciteSetBstMidEndSepPunct{\mcitedefaultmidpunct}
{\mcitedefaultendpunct}{\mcitedefaultseppunct}\relax
\EndOfBibitem
\bibitem[Akram \latin{et~al.}(2021)Akram, LaBollita, Dey, Kapeghian, Erten, and
  Botana]{akram2021moire}
Akram,~M.; LaBollita,~H.; Dey,~D.; Kapeghian,~J.; Erten,~O.; Botana,~A.~S.
  Moir{\'e} skyrmions and chiral magnetic phases in twisted CrX3 (X= I, Br, and
  Cl) bilayers. \emph{Nano Letters} \textbf{2021}, \emph{21}, 6633--6639\relax
\mciteBstWouldAddEndPuncttrue
\mciteSetBstMidEndSepPunct{\mcitedefaultmidpunct}
{\mcitedefaultendpunct}{\mcitedefaultseppunct}\relax
\EndOfBibitem
\bibitem[Zheng(2023)]{6}
Zheng,~F. Magnetic Skyrmion Lattices in a Novel 2D-Twisted Bilayer Magnet.
  \emph{Advanced Functional Materials} \textbf{2023}, \emph{33}, 2206923\relax
\mciteBstWouldAddEndPuncttrue
\mciteSetBstMidEndSepPunct{\mcitedefaultmidpunct}
{\mcitedefaultendpunct}{\mcitedefaultseppunct}\relax
\EndOfBibitem
\bibitem[Li \latin{et~al.}(2018)Li, Wang, Guo, Gu, Sun, He, Zhou, Gu, Nie, and
  Pan]{27}
Li,~Y.~F.; Wang,~W.; Guo,~W.; Gu,~C.~Y.; Sun,~H.~Y.; He,~L.; Zhou,~J.;
  Gu,~Z.~B.; Nie,~Y.~F.; Pan,~X.~Q. Electronic structure of ferromagnetic
  semiconductor ${\mathrm{CrGeTe}}_{3}$ by angle-resolved photoemission
  spectroscopy. \emph{Phys. Rev. B} \textbf{2018}, \emph{98}, 125127\relax
\mciteBstWouldAddEndPuncttrue
\mciteSetBstMidEndSepPunct{\mcitedefaultmidpunct}
{\mcitedefaultendpunct}{\mcitedefaultseppunct}\relax
\EndOfBibitem
\bibitem[Webster and Yan(2018)Webster, and Yan]{11}
Webster,~L.; Yan,~J.-A. Strain-tunable magnetic anisotropy in monolayer
  ${\mathrm{CrCl}}_{3}$, ${\mathrm{CrBr}}_{3}$, and ${\mathrm{CrI}}_{3}$.
  \emph{Phys. Rev. B} \textbf{2018}, \emph{98}, 144411\relax
\mciteBstWouldAddEndPuncttrue
\mciteSetBstMidEndSepPunct{\mcitedefaultmidpunct}
{\mcitedefaultendpunct}{\mcitedefaultseppunct}\relax
\EndOfBibitem
\bibitem[Zheng \latin{et~al.}(2018)Zheng, Zhao, Liu, Li, Zhou, Zhang, and
  Zhang]{26}
Zheng,~F.; Zhao,~J.; Liu,~Z.; Li,~M.; Zhou,~M.; Zhang,~S.; Zhang,~P. Tunable
  spin states in the two-dimensional magnet CrI$_3$. \emph{Nanoscale}
  \textbf{2018}, \emph{10}, 14298--14303\relax
\mciteBstWouldAddEndPuncttrue
\mciteSetBstMidEndSepPunct{\mcitedefaultmidpunct}
{\mcitedefaultendpunct}{\mcitedefaultseppunct}\relax
\EndOfBibitem
\bibitem[Verzhbitskiy \latin{et~al.}(2020)Verzhbitskiy, Kurebayashi, Cheng,
  Zhou, Khan, Feng, and Eda]{10}
Verzhbitskiy,~I.~A.; Kurebayashi,~H.; Cheng,~H.; Zhou,~J.; Khan,~S.;
  Feng,~Y.~P.; Eda,~G. Controlling the magnetic anisotropy in
  Cr$_2$Ge$_2$Te$_6$ by electrostatic gating. \emph{Nature Electronics}
  \textbf{2020}, \emph{3}, 460--465\relax
\mciteBstWouldAddEndPuncttrue
\mciteSetBstMidEndSepPunct{\mcitedefaultmidpunct}
{\mcitedefaultendpunct}{\mcitedefaultseppunct}\relax
\EndOfBibitem
\bibitem[Jiang \latin{et~al.}(2018)Jiang, Li, Wang, Mak, and Shan]{7}
Jiang,~S.; Li,~L.; Wang,~Z.; Mak,~K.~F.; Shan,~J. Controlling magnetism in 2D
  CrI$_3$ by electrostatic doping. \emph{Nature Nanotechnology} \textbf{2018},
  \emph{13}, 549--553\relax
\mciteBstWouldAddEndPuncttrue
\mciteSetBstMidEndSepPunct{\mcitedefaultmidpunct}
{\mcitedefaultendpunct}{\mcitedefaultseppunct}\relax
\EndOfBibitem
\bibitem[Zhang \latin{et~al.}(2015)Zhang, Qu, Zhu, and Lam]{C5TC02840J}
Zhang,~W.-B.; Qu,~Q.; Zhu,~P.; Lam,~C.-H. Robust intrinsic ferromagnetism and
  half semiconductivity in stable two-dimensional single-layer chromium
  trihalides. \emph{J. Mater. Chem. C} \textbf{2015}, \emph{3},
  12457--12468\relax
\mciteBstWouldAddEndPuncttrue
\mciteSetBstMidEndSepPunct{\mcitedefaultmidpunct}
{\mcitedefaultendpunct}{\mcitedefaultseppunct}\relax
\EndOfBibitem
\bibitem[Hou \latin{et~al.}()Hou, Wei, Yang, Wang, Ren, and Zhang]{Hou2023}
Hou,~Y.; Wei,~Y.; Yang,~D.; Wang,~K.; Ren,~K.; Zhang,~G. Enhancing the Curie
  Temperature in Cr$_2$Ge$_2$Te$_6$ via Charge Doping: A First-Principles
  Study. \emph{Molecules} \textbf{2023}, \emph{28}, 3893\relax
\mciteBstWouldAddEndPunctfalse
\mciteSetBstMidEndSepPunct{\mcitedefaultmidpunct}
{}{\mcitedefaultseppunct}\relax
\EndOfBibitem
\bibitem[Deng \latin{et~al.}(2018)Deng, Yu, Song, Zhang, Wang, Sun, Yi, Wu, Wu,
  Zhu, Wang, Chen, and Zhang]{12}
Deng,~Y.; Yu,~Y.; Song,~Y.; Zhang,~J.; Wang,~N.~Z.; Sun,~Z.; Yi,~Y.; Wu,~Y.~Z.;
  Wu,~S.; Zhu,~J.; Wang,~J.; Chen,~X.~H.; Zhang,~Y. Gate-tunable
  room-temperature ferromagnetism in two-dimensional Fe$_3$GeTe$_2$.
  \emph{Nature} \textbf{2018}, \emph{563}, 94--99\relax
\mciteBstWouldAddEndPuncttrue
\mciteSetBstMidEndSepPunct{\mcitedefaultmidpunct}
{\mcitedefaultendpunct}{\mcitedefaultseppunct}\relax
\EndOfBibitem
\bibitem[Fei \latin{et~al.}(2018)Fei, Huang, Malinowski, Wang, Song, Sanchez,
  Yao, Xiao, Zhu, May, Wu, Cobden, Chu, and Xu]{13}
Fei,~Z.; Huang,~B.; Malinowski,~P.; Wang,~W.; Song,~T.; Sanchez,~J.; Yao,~W.;
  Xiao,~D.; Zhu,~X.; May,~A.~F.; Wu,~W.; Cobden,~D.~H.; Chu,~J.-H.; Xu,~X.
  Two-dimensional itinerant ferromagnetism in atomically thin Fe$_3$GeTe$_2$.
  \emph{Nature Materials} \textbf{2018}, \emph{17}, 778--782\relax
\mciteBstWouldAddEndPuncttrue
\mciteSetBstMidEndSepPunct{\mcitedefaultmidpunct}
{\mcitedefaultendpunct}{\mcitedefaultseppunct}\relax
\EndOfBibitem
\bibitem[Li \latin{et~al.}(2020)Li, Xia, Su, Yu, Fu, Chen, Wang, Yu, Zou, and
  Guo]{Li2020}
Li,~Z.; Xia,~W.; Su,~H.; Yu,~Z.; Fu,~Y.; Chen,~L.; Wang,~X.; Yu,~N.; Zou,~Z.;
  Guo,~Y. Magnetic critical behavior of the van der Waals Fe$_5$GeTe$_2$
  crystal with near room temperature ferromagnetism. \emph{Scientific Reports}
  \textbf{2020}, \emph{10}, 15345\relax
\mciteBstWouldAddEndPuncttrue
\mciteSetBstMidEndSepPunct{\mcitedefaultmidpunct}
{\mcitedefaultendpunct}{\mcitedefaultseppunct}\relax
\EndOfBibitem
\bibitem[Bonilla \latin{et~al.}(2018)Bonilla, Kolekar, Ma, Diaz, Kalappattil,
  Das, Eggers, Gutierrez, Phan, and Batzill]{15}
Bonilla,~M.; Kolekar,~S.; Ma,~Y.; Diaz,~H.~C.; Kalappattil,~V.; Das,~R.;
  Eggers,~T.; Gutierrez,~H.~R.; Phan,~M.-H.; Batzill,~M. Strong
  room-temperature ferromagnetism in VSe2 monolayers on van der Waals
  substrates. \emph{Nature Nanotechnology} \textbf{2018}, \emph{13},
  289--293\relax
\mciteBstWouldAddEndPuncttrue
\mciteSetBstMidEndSepPunct{\mcitedefaultmidpunct}
{\mcitedefaultendpunct}{\mcitedefaultseppunct}\relax
\EndOfBibitem
\bibitem[Lin \latin{et~al.}(2016)Lin, Zhuang, Yan, Ward, Puretzky, Rouleau,
  Gai, Liang, Meunier, Sumpter, Ganesh, Kent, Geohegan, Mandrus, and
  Xiao]{CrSiTe}
Lin,~M.-W.; Zhuang,~H.~L.; Yan,~J.; Ward,~T.~Z.; Puretzky,~A.~A.;
  Rouleau,~C.~M.; Gai,~Z.; Liang,~L.; Meunier,~V.; Sumpter,~B.~G.; Ganesh,~P.;
  Kent,~P. R.~C.; Geohegan,~D.~B.; Mandrus,~D.~G.; Xiao,~K. Ultrathin
  nanosheets of CrSiTe$_3$: a semiconducting two-dimensional ferromagnetic
  material. \emph{J. Mater. Chem. C} \textbf{2016}, \emph{4}, 315--322\relax
\mciteBstWouldAddEndPuncttrue
\mciteSetBstMidEndSepPunct{\mcitedefaultmidpunct}
{\mcitedefaultendpunct}{\mcitedefaultseppunct}\relax
\EndOfBibitem
\bibitem[Zhang \latin{et~al.}(2022)Zhang, Guo, Wu, Wen, Yang, Jin, Zhang, and
  Chang]{14}
Zhang,~G.; Guo,~F.; Wu,~H.; Wen,~X.; Yang,~L.; Jin,~W.; Zhang,~W.; Chang,~H.
  Above-room-temperature strong intrinsic ferromagnetism in 2D van der Waals
  Fe$_3$GaTe$_2$ with large perpendicular magnetic anisotropy. \emph{Nature
  Communications} \textbf{2022}, \emph{13}, 5067\relax
\mciteBstWouldAddEndPuncttrue
\mciteSetBstMidEndSepPunct{\mcitedefaultmidpunct}
{\mcitedefaultendpunct}{\mcitedefaultseppunct}\relax
\EndOfBibitem
\bibitem[Yin \latin{et~al.}(2023)Yin, Zhang, Jin, Di, Wu, Zhang, Zhang, and
  Chang]{20}
Yin,~H.; Zhang,~P.; Jin,~W.; Di,~B.; Wu,~H.; Zhang,~G.; Zhang,~W.; Chang,~H.
  Fe$_3$GaTe$_2$/MoSe$_2$ ferromagnet/semiconductor 2D van der Waals
  heterojunction for room-temperature spin-valve devices. \emph{CrystEngComm}
  \textbf{2023}, \emph{25}, 1339--1346\relax
\mciteBstWouldAddEndPuncttrue
\mciteSetBstMidEndSepPunct{\mcitedefaultmidpunct}
{\mcitedefaultendpunct}{\mcitedefaultseppunct}\relax
\EndOfBibitem
\bibitem[Jin \latin{et~al.}(2023)Jin, Zhang, Wu, Yang, Zhang, and Chang]{22}
Jin,~W.; Zhang,~G.; Wu,~H.; Yang,~L.; Zhang,~W.; Chang,~H. Room-temperature
  spin-valve devices based on Fe$_3$GaTe$_2$/MoS$_2$/Fe$_3$GaTe$_2$ 2D van der
  Waals heterojunctions. \emph{Nanoscale} \textbf{2023}, \emph{15},
  5371--5378\relax
\mciteBstWouldAddEndPuncttrue
\mciteSetBstMidEndSepPunct{\mcitedefaultmidpunct}
{\mcitedefaultendpunct}{\mcitedefaultseppunct}\relax
\EndOfBibitem
\bibitem[Jin \latin{et~al.}(2023)Jin, Zhang, Wu, Yang, Zhang, and Chang]{23}
Jin,~W.; Zhang,~G.; Wu,~H.; Yang,~L.; Zhang,~W.; Chang,~H. Room-Temperature and
  Tunable Tunneling Magnetoresistance in Fe3GaTe2-Based 2D van der Waals
  Heterojunctions. \emph{ACS Applied Materials {\&} Interfaces} \textbf{2023},
  \emph{15}, 36519--36526\relax
\mciteBstWouldAddEndPuncttrue
\mciteSetBstMidEndSepPunct{\mcitedefaultmidpunct}
{\mcitedefaultendpunct}{\mcitedefaultseppunct}\relax
\EndOfBibitem
\bibitem[Wang \latin{et~al.}(2023)Wang, Wang, Xie, Zhang, Wu, Zhou, Zhu, Ning,
  Wang, Tan, Wang, Du, Zhao, Chang, Zheng, Geng, and Tian]{24}
Wang,~C. \latin{et~al.}  Sign-tunable exchange bias effect in
  proton-intercalated ${\mathrm{Fe}}_{3}\mathrm{Ga}{\mathrm{Te}}_{2}$
  nanoflakes. \emph{Phys. Rev. B} \textbf{2023}, \emph{107}, L140409\relax
\mciteBstWouldAddEndPuncttrue
\mciteSetBstMidEndSepPunct{\mcitedefaultmidpunct}
{\mcitedefaultendpunct}{\mcitedefaultseppunct}\relax
\EndOfBibitem
\bibitem[Kresse and Hafner(1993)Kresse, and Hafner]{30}
Kresse,~G.; Hafner,~J. Ab initio molecular dynamics for liquid metals.
  \emph{Phys. Rev. B} \textbf{1993}, \emph{47}, 558--561\relax
\mciteBstWouldAddEndPuncttrue
\mciteSetBstMidEndSepPunct{\mcitedefaultmidpunct}
{\mcitedefaultendpunct}{\mcitedefaultseppunct}\relax
\EndOfBibitem
\bibitem[Kresse and Furthm\"uller(1996)Kresse, and Furthm\"uller]{31}
Kresse,~G.; Furthm\"uller,~J. Efficient iterative schemes for ab initio
  total-energy calculations using a plane-wave basis set. \emph{Phys. Rev. B}
  \textbf{1996}, \emph{54}, 11169--11186\relax
\mciteBstWouldAddEndPuncttrue
\mciteSetBstMidEndSepPunct{\mcitedefaultmidpunct}
{\mcitedefaultendpunct}{\mcitedefaultseppunct}\relax
\EndOfBibitem
\bibitem[Bl\"ochl(1994)]{33}
Bl\"ochl,~P.~E. Projector augmented-wave method. \emph{Phys. Rev. B}
  \textbf{1994}, \emph{50}, 17953--17979\relax
\mciteBstWouldAddEndPuncttrue
\mciteSetBstMidEndSepPunct{\mcitedefaultmidpunct}
{\mcitedefaultendpunct}{\mcitedefaultseppunct}\relax
\EndOfBibitem
\bibitem[Perdew \latin{et~al.}(1996)Perdew, Burke, and Ernzerhof]{32}
Perdew,~J.~P.; Burke,~K.; Ernzerhof,~M. Generalized Gradient Approximation Made
  Simple. \emph{Phys. Rev. Lett.} \textbf{1996}, \emph{77}, 3865--3868\relax
\mciteBstWouldAddEndPuncttrue
\mciteSetBstMidEndSepPunct{\mcitedefaultmidpunct}
{\mcitedefaultendpunct}{\mcitedefaultseppunct}\relax
\EndOfBibitem
\bibitem[Togo and Tanaka(2015)Togo, and Tanaka]{34}
Togo,~A.; Tanaka,~I. First principles phonon calculations in materials science.
  \emph{Scripta Materialia} \textbf{2015}, \emph{108}, 1--5\relax
\mciteBstWouldAddEndPuncttrue
\mciteSetBstMidEndSepPunct{\mcitedefaultmidpunct}
{\mcitedefaultendpunct}{\mcitedefaultseppunct}\relax
\EndOfBibitem
\bibitem[Metropolis and Ulam(1949)Metropolis, and Ulam]{36}
Metropolis,~N.; Ulam,~S. The Monte Carlo Method. \emph{Journal of the American
  Statistical Association} \textbf{1949}, \emph{44}, 335--341, PMID:
  18139350\relax
\mciteBstWouldAddEndPuncttrue
\mciteSetBstMidEndSepPunct{\mcitedefaultmidpunct}
{\mcitedefaultendpunct}{\mcitedefaultseppunct}\relax
\EndOfBibitem
\bibitem[Xiang \latin{et~al.}(2011)Xiang, Kan, Wei, Whangbo, and Gong]{35}
Xiang,~H.~J.; Kan,~E.~J.; Wei,~S.-H.; Whangbo,~M.-H.; Gong,~X.~G. Predicting
  the spin-lattice order of frustrated systems from first principles.
  \emph{Phys. Rev. B} \textbf{2011}, \emph{84}, 224429\relax
\mciteBstWouldAddEndPuncttrue
\mciteSetBstMidEndSepPunct{\mcitedefaultmidpunct}
{\mcitedefaultendpunct}{\mcitedefaultseppunct}\relax
\EndOfBibitem
\bibitem[Yu and Trinkle(2011)Yu, and Trinkle]{bader}
Yu,~M.; Trinkle,~D.~R. {Accurate and efficient algorithm for Bader charge
  integration}. \emph{The Journal of Chemical Physics} \textbf{2011},
  \emph{134}, 064111\relax
\mciteBstWouldAddEndPuncttrue
\mciteSetBstMidEndSepPunct{\mcitedefaultmidpunct}
{\mcitedefaultendpunct}{\mcitedefaultseppunct}\relax
\EndOfBibitem
\end{mcitethebibliography}
\end{document}